\documentclass[prl,preprint,endfloats]{revtex4}

\usepackage[dvipdfmx]{graphicx}
\usepackage{bm}
\usepackage{pifont}
\usepackage{amsmath}
\usepackage{color}
\usepackage{comment}
\usepackage{transparent}
\usepackage{textcomp}

\begin{document}

\title{Large magneto-optical Kerr effect induced by collinear antiferromagnetic order}

\author{H. Yoshimochi$^{1,*}$, K. Yoshida$^{1,*}$, R. Oiwa$^{2,3}$, T. Nomoto$^{4}$, N. D. Khanh$^{1}$, A. Kitaori$^{1,5,6}$, R. Takagi$^{5,7}$, R. Arita$^{2, 8}$, S. Seki$^{1,6,9}$}

\affiliation{$^1$ Department of Applied Physics and Quantum-Phase Electronics Center (QPEC), University of Tokyo, Tokyo 113-8656, Japan, \\ 
$^2$ RIKEN Center for Emergent Matter Science (CEMS), Wako 351-0198, Japan,\\
$^3$ Department of Physics, Hokkaido University, Sapporo 060-0810, Japan, \\
$^4$ Department of Physics, Tokyo Metropolitan University, Hachioji, Tokyo 192-0397, Japan,\\
$^5$ PRESTO, Japan Science and Technology Agency (JST), Kawaguchi 332-0012, Japan, \\ 
$^6$ Research Center for Advanced Science and Technology (RCAST), University of Tokyo, Tokyo 153-8904, Japan, \\ 
$^7$ Institute for Solid State Physics, University of Tokyo, Kashiwa 277-8581, Japan,\\
$^8$ Department of Physics, University of Tokyo, Tokyo 113-8656, Japan, \\ 
$^9$ Institute of Engineering Innovation, University of Tokyo, Tokyo 113-0032, Japan\\ $*$ These authors contributed equally.}

\begin{abstract}
{\bf In modern technology, the optical readout of magnetic information is conventionally achieved by the magneto-optical Kerr effect, i.e.,  the polarization rotation of reflected light. The Kerr rotation is sensitive to time-reversal symmetry breaking and generally proportional to magnetization, enabling optical readout of the $\uparrow$ and $\downarrow$ spin states in ferromagnets. By contrast, antiferromagnets with a collinear antiparallel spin arrangement have long been considered inactive to such magneto-optical responses, because of $\mathcal{T}t$ symmetry (time-reversal $\mathcal{T}$ followed by translation $t$ symmetry) and lack of macroscopic magnetization. Here, we report the observation of giant magneto-optical Kerr effect in a room-temperature antiferromagnetic insulator $\alpha$-Fe$_2$O$_3$. In this compound, the $\uparrow \downarrow$ and $ \downarrow \uparrow$ spin states induce the opposite sign of spontaneous Kerr effect, whose Kerr rotation angle turned out to be exceptionally large ($\theta_{\rm K} \sim 80$ mdeg, comparable to typical ferromagnets). Our first-principles calculations successfully reproduce both the absolute magnitude and spectral shape of the Kerr rotation and ellipticity with remarkable accuracy, which unambiguously proves that it originates from a $\mathcal{T}t$-symmetry-broken collinear antiferromagnetic order, rather than magnetization. This compound hosts temperature-dependent transition between easy-plane and easy-axis antiferromagnetic states, and their contrasting behaviors are also investigated in detail. The present results demonstrate that even a simple collinear antiferromagnetic order can induce a giant magneto-optical Kerr effect, and highlight $\mathcal{T}t$-symmetry-broken antiferromagnets as a promising material platform for highly sensitive optical detection of $\uparrow\downarrow$ and $\downarrow\uparrow$ spin states.
}
\end{abstract}

\maketitle

Currently, magnetic information is mainly stored by the $\uparrow$ and $\downarrow$ spin states of ferromagnets. These two states are related by the time-reversal operation $\mathcal{T}$, and their distinguishability originates from the breaking of time-reversal symmetry in ferromagnets (Left panel of Fig. 1a). Such a time-reversal symmetry breaking gives rise to various functional off-diagonal responses, such as spontaneous Hall effect\cite{RMP_AHE} and the spontaneous magneto-optical Kerr effect\cite{KerrTextbook}, both of which are frequently used for electrical and optical readout of magnetic information.

Here, the spontaneous Hall effect represents the generation of transverse electric field ${\bf E}$ $(\parallel y)$ by the application of electric current ${\bf j}$ $(\parallel x)$, which originates from the off-diagonal component ($\sigma_{xy}$) of the electrical conductivity tensor $\tilde{\sigma}$ defined by $\bf{j}=\tilde{\sigma}\bf{E}$. In ferromagnets, $\sigma_{xy}$ is generally proportional to magnetization $M$ along the $z$-direction ($M_z$)\cite{RMP_AHE}. On the other hand, the magneto-optical Kerr effect refers to the change in the polarization state of reflected light, when linearly polarized light is incident on the surface of a magnetic material (Left panel of Fig. 1b)\cite{KerrTextbook}. This effect is characterized by the Kerr rotation angle $\theta_{\rm K}$, describing the rotation of the polarization axis, and the Kerr ellipticity $\eta_{\rm K}$, describing the induced ellipticity. The magneto-optical Kerr effect can be regarded as an optical-frequency extension of the Hall effect. For example, in the polar Kerr geometry, where incident direction of light ($z$) is normal to the sample surface, the relation $(\theta_{\rm K}+i\eta_{\rm K}) \propto \sigma_{xy}$ holds. Consequently, $\theta_{\rm K}$ and $\eta_{\rm K}$ are also expected to be proportional to $M_z$ in ferromagnets. These phenomena provide electrical and optical probes of the $\uparrow$ and $\downarrow$ spin states in ferromagnets.

In contrast, it has long been considered that such functional responses cannot be expected in antiferromagnets (AFMs) with collinear antiparallel spin alignment, because of their vanishing net magnetization. This can be intuitively understood from symmetry considerations. In conventional collinear antiferromagnets, one can consider the $\uparrow\downarrow$ and $\downarrow\uparrow$ spin configurations, which are converted into each other by the time-reversal operation $\mathcal{T}$. As illustrated in the middle panel of Fig.~1a, however, these two configurations are identical under a lattice translation $t$ and are therefore indistinguishable. As a result, the combined $\mathcal{T}t$ symmetry (TtS) is preserved, prohibiting the appearance of aforementioned off-diagonal responses.

Very recently, however, it has been theoretically pointed out that even a collinear antiferromagnetic order can break TtS and induce spontaneous Hall/Kerr effect when combined with a crystal lattice of appropriate symmetry \cite{CrystalHall, Naka}. For example, when non-magnetic ions are arranged in a staggered manner, as shown in the right panel of Fig.~1a, the $\uparrow\downarrow$ and $\downarrow\uparrow$ states are no longer equivalent under a translation. In this case, TtS is broken, and the emergence of a spontaneous Hall/Kerr effect is often allowed even in the absence of net magnetization (Right panel of Fig.~1b). This new class of collinear antiferromagnets with $\mathcal{T}t$ symmetry breaking, often termed "altermagnets", is predicted to exhibit ferromagnet-like functional responses\cite{Altermagnet, AltermagnetReview, Hayami_SpinSplit, Naka_Organic, MnTe_ARPES, MnTe_XMCD, MnTe_XMCD_Imaging}, while offering unique advantages such as negligibly small stray fields and ultrafast spin dynamics \cite{RMP_Ono, AFSpintronicsReview}.

According to the latest experimental studies, such a spontaneous Hall effect driven by antiferromagnetic order (rather than $M$) has indeed been identified in a few TtS-broken collinear antiferromagnets such as FeS\cite{FeS_Hall} and MnTe\cite{MnTe_Hall}. In this case, the Berry curvature of the electronic band structure acts as a fictitious magnetic field for conduction electrons, resulting in the spontaneous Hall response without the necessity of $M$. These findings suggest that TtS-broken antiferromagnets may also host a spontaneous magneto-optical Kerr effect independently of magnetization, since it follows the same symmetry-based selection rules as the Hall effect. Importantly, while the Hall effect can only be observed in metals, the magneto-optical Kerr effect is accessible even in insulating materials. Moreover, its spectral information will provide a better means to understand the microscopic origin of the observed response.

In this work, we report the observation of giant magneto-optical Kerr effect in a room-temperature antiferromagnetic insulator $\alpha$-Fe$_2$O$_3$, which turned out to originate from a TtS-broken collinear antiferromagnetic order, rather than net magnetization. Our first-principles calculations successfully reproduce both the absolute magnitude and spectral shape of the Kerr rotation and ellipticity with remarkable accuracy. This compound hosts temperature-dependent transition between easy-plane and easy-axis antiferromagnetic states, and their contrasting behaviors are also investigated in detail. The present results demonstrate that even a simple collinear antiferromagnetic order can induce a giant magneto-optical Kerr effect, and highlight TtS-broken antiferromagnets as a promising material platform for highly sensitive optical detection of $\uparrow\downarrow$ and $\downarrow\uparrow$ spin states.

Our target material $\alpha$-Fe$_2$O$_3$ crystallizes in the corundum structure with the trigonal space group $R\bar{3}c$, as shown in Fig. 2a. The crystal lattice consists of alternating stacking of Fe and O ion layers along the $c$-axis, and a crystallographic unit cell contains four magnetic Fe$^{3+}$ ($S=5/2$) sites. Each Fe ion is sandwiched by a pair of oppositely oriented triangular plaquettes of O ions, which realizes the staggered manner of non-magnetic ion distribution as considered in the right panel of Fig. 1a. Magnetic interaction is antiferromagnetic between the nearest neighbor Fe layers, and ferromagnetic between the second nearest neighbor layers. As a result, the easy-plane collinear antiferromagnetic order as shown in Fig. 2a and the middle panel of Fig. 3g is stabilized below the magnetic ordering temperature $T_{\rm N}\sim 948$ K, where antiparallel spins are oriented along the $y$-axis (i.e., the direction normal to both $a$- and $c$-axes)\cite{Fe2O3_Shibata}. 

In $\alpha$-Fe$_2$O$_3$, the space-inversion symmetry is broken at the middle point between two neighboring Fe sites ($i$ and $j$), which leads to the appearance of Dzyaloshinskii-Moriya (DM) interaction $\mathcal{H}_{\rm DM} = \sum_{ij} {\bf D}^c_{ij} \cdot ({\bf S}_i \times {\bf S}_j)$\cite{Fe2O3_Moriya}. Here, ${\bf S}_i$ and ${\bf S}_j$ are spins at the sites $i$ and $j$, and ${\bf D}^c_{ij}$ is a DM vector parallel to the $c$-axis. To obtain the energy gain by DM interaction, spins are slightly tilted from the $y$-axis toward the $a$-axis, leading to the emergence of tiny spontaneous magnetization $\Delta M_{\rm cant}$ along the $a$-axis. Notably, this compound is also known to host the Morin transition at  $T_{\rm M} \sim 260$ K, where the sign of magnetic anisotropy changes\cite{Morin}. Below $T_{\rm M}$, easy-axis antiferromagnetic state is realized, in which antiparallel spins are oriented along the $c$-axis as shown in the left panel of Fig. 3g. In the latter phase, the aforementioned DM-induced spin tilting and associated spontaneous magnetization are absent. Previously, the presence of finite magneto-optical Kerr responses in $\alpha$-Fe$_2$O$_3$ has been reported\cite{Fe2O3_Kerr1, Fe2O3_Kerr2, Fe2O3_Kerr3}. Nevertheless, the clarification of its detailed microscopic mechanism (i.e., theoretical explanation of absolute magnitude and spectral shape of $\theta_{\rm K}$ and $\eta_{\rm K}$), as well as the relationship between two distinctive AFM orders, remains an essential challenge.

In the following, we focus on the polar Kerr geometry, where the incident and reflected lights are normal to the $a$-plane surface of the sample. Figures 2b and c display the magnetic field dependence of magnetization $M$ and Kerr rotation angle $\theta_{\rm K}$, measured at 300 K (i.e., the easy-plane antiferromagnetic state) with $B \parallel a$. In both $M$ and $\theta_{\rm K}$ profiles, a clear step-like anomaly is observed around $B=0$. Despite very small amplitude of spontaneous magnetization $\Delta M_{\rm cant} \sim 0.006 \mu_{\rm B}$/Fe, the observed spontaneous Kerr rotation angle ($\Delta \theta_{\rm K} \sim 80$ mdeg) is exceptionally large, which is comparable or even larger than typical ferromagnets. Importantly, while $M$ increases linearly with a magnetic field, $\theta_{\rm K}$ remains nearly constant up to 5 T. These features imply that magnetization $M$ is not the main source of the observed $\theta_{\rm K}$ signal. (We also confirmed that Kerr ellipticity $\eta_{\rm K}$ shows a similar $B$-dependence as $\theta_{\rm K}$ as detailed in Supplementary Note I and Supplementary Fig. 1.)

Such a spontaneous Kerr signal in the easy-plane antiferromagnetic phase can be naturally understood by considering its magnetic symmetry. According to the Neumann's principle, when the magnetic point group of the target system is specified, the symmetry-imposed shape of optical conductivity tensor $\tilde {\sigma}$ (i.e., the presence/absence of each tensor component) is uniquely identified\cite{Conductivity_Symmetry, Conductivity_Symmetry2}. Figure 3g summarizes the magnetic point group and the corresponding symmetry-imposed shape of the conductivity tensor for various collinear spin configurations on a $\alpha$-Fe$_2$O$_3$ crystal lattice. Importantly, the magnetic point group of the easy-plane antiferromagnetic state is $2/m$, which is the same as the one for the ferromagnetic state with $M \parallel a$. As a result, these two states are characterized by the common form of the optical conductivity tensor with non-zero off-diagonal component $\sigma_{yz}$\cite{Conductivity_Symmetry, Conductivity_Symmetry2}. This means that reflected light in the easy-plane antiferromagnetic state should behave as if there is fictitious magnetic field (or fictitious magnetization) along the $x$ ($|| a$) axis, which leads to the appearance of a spontaneous polar Kerr effect for the $a$-plane surface. In such an environment, DM interaction is allowed to cause additional canting of antiferromagnetically aligned moments toward the $a$-direction, which explains the appearance of a weak spontaneous magnetization $\Delta M_{\rm cant} \parallel a$\cite{Fe2O3_Moriya}. Note that the aforementioned spontaneous Kerr effect is allowed even without this additional spin canting, since the easy-plane collinear antiferromagnetic order itself can break TtS and host non-zero $\sigma_{yz}$. There are two degenerated time-reversal domains, i.e., the domains A and B characterized by the $\uparrow \downarrow$ and $\downarrow \uparrow$ spin arrangements, respectively (Figs. 2a and d). They are converted into each other by the time-reversal operation, and therefore should be characterized by opposite signs of the fictitious magnetic field, spontaneous Kerr rotation $\Delta \theta_{\rm K}$ and $\Delta M_{\rm cant}$. By applying the positive (negative) sign of external magnetic field along the $a$-axis, their degeneracy is lifted through $\Delta M_{\rm cant}$ and the domain A (domain B) can be selected. This suggests that the step-like anomaly observed in the $\theta_{\rm K}$ profile should represent the $B$-induced switching between the domains A and B.

Below the Morin transition temperature $T_{\rm M}$, the sign of magnetic anisotropy is changed and the easy-axis antiferromagnetic state is realized. As shown in the left panel of Fig. 3g, the corresponding magnetic point group is $\bar{3}'m$, which indicates the absence of any off-diagonal component of the conductivity tensor and associated spontaneous Kerr effect. To confirm this symmetry-based prediction, magnetic-field-dependence of $M$ and $\theta_{\rm K}$ are measured at $50$~K in the easy-axis antiferromagnetic phase (Figs. 3a and b). Unlike the easy-plane antiferromagnetic phase, the step-like anomalies around $B=0$ associated with spontaneous $\Delta M_{\rm cant}$ and $\Delta \theta_{\rm K}$ are absent, which is consistent with the aforementioned symmetry analysis. 

For better understanding of the contrasting behaviors between the easy-plane and easy-axis AFM states, similar measurements are also performed at 245 K (Figs. 3c and d) close to the Morin transition temperature $T_{\rm M} \sim 260$ K. Here, the application of $B \parallel a$ induces a transition from easy-axis AFM state to easy-plane AFM state at 2.5 T, which is identified as a small anomaly in the magnetization profile. In the easy-plane AFM state, $\theta_{\rm K}$ remains nearly constant against $B$. It indicates that the observed Kerr signal is governed by the antiferromagnetic order parameter, and the $M$-linear component of $\theta_{\rm K}$ is negligibly small. In the easy-axis AFM state, on the other hand, the linear slope of $\theta_{\rm K}$-$B$ curve is much larger than easy-plane AFM phase, despite similar amplitude of linear slopes in $M$-$B$ profiles between these two phases. It suggests that $B$-linear increase of $\theta_{\rm K}$ in the easy-axis AFM phase is probably not caused by $M$, but rather by the $B$-induced tilting of N\'{e}el vector (defined by ${\bf L}={\bf S}_{1} - {\bf S}_{2}$, with ${\bf S}_{1}$ and ${\bf S}_{2}$ representing the spins at two distinctive sublattices) away from the $c$-axis towards the $y$-axis. Since this ${\bf L}$-tilting induces the same spin component as the easy-plane AFM state, it leads to the appearance of finite $\theta_{\rm K}$ scaling with the $B$-induced {\bf L}-tilting angle. This direction of ${\bf L}$-tilting should be favored under ${\bf B} \parallel a$, because the combination of ${\bf M} (={\bf S}_{1} + {\bf S}_{2}) \parallel a$ and ${\bf L} \parallel y$ component leads to the additional DM energy gain ${\bf D}^c_{12} \cdot ({\bf S}_1 \times {\bf S}_2)$ (which is the source of weak spontaneous ${\bf M} \parallel a$ in the easy-plane AFM state with ${\bf L} \parallel y$). In this context, the observed $B$-linear increase of $\theta_{\rm K}$ in the easy-axis AFM state can be also ascribed to the $B$-induced component of easy-plane-type antiferromagnetic order parameter, rather than small net magnetization.

Figure 3e summarizes the experimentally measured amplitude of $\theta_{\rm K}$ plotted on the $B$-$T$ phase diagram. The easy-plane antiferromagnetic state is always characterized by large amplitude of $\theta_{\rm K}$, which remains almost constant against $B$. In the easy-axis antiferromagnetic state, on the other hand, $\theta_{\rm K}$ linearly increases as a function of $B$, while its amplitude is smaller. Figure 3f indicates the temperature dependence of spontaneous Kerr rotation angle $\Delta \theta_{\rm K}$, obtained by linearly extrapolating the $\theta_{\rm K}$-$B$ curve to $B=0$. It clearly demonstrates that large spontaneous Kerr effect is observable only in the easy-plane antiferromagnetic state, in accord with the symmetry-based prediction in Fig. 3g.

To understand the microscopic origin of the observed giant Kerr response, we examined the spectroscopic features of the magneto-optical signals. Figure 4b displays the experimental energy spectra of the Kerr rotation $\theta_{\rm K}$ and ellipticity $\eta_{\rm K}$ measured at 300 K (easy-plane AFM phase). The $\theta_{\rm K}$ spectrum exhibits a broad peak at $\sim 2.3$ eV, accompanied by a sign change in the $\eta_{\rm K}$ spectrum. This dispersive line shape in $\eta_{\rm K}$ is the Kramers-Kronig counterpart to the absorptive peak structure observed in the $\theta_{\rm K}$ spectrum, which validates the resonance nature of the electronic transitions responsible for the giant Kerr signal.

To theoretically account for these features, we performed first-principles calculations based on density functional theory (DFT). Figure 4a shows the calculated electronic band structure and density of states (DOS) in the easy-plane AFM state. The DOS analysis reveals that the Fe $3d$ orbitals are split into occupied and unoccupied bands, separated by the O $2p$ bands. This electronic configuration characterizes $\alpha$-Fe$_2$O$_3$ as a charge-transfer insulator, where the energy gap of around 2 eV corresponds to the optical transition from the occupied O $2p$ orbitals to the unoccupied Fe $3d$ orbitals.

Figure 4c presents the Kerr spectra theoretically calculated for the easy-plane AFM state with $M=0$, based on the electronic structure obtained by DFT calculation. The calculated spectra exhibit a broad peak structure in $\theta_{\rm K}$ at $\sim 2.5$ eV, accompanied by a sign reversal in $\eta_{\rm K}$. These features are in good agreement with the experimental data in Fig. 4b. In both experimental and theoretical data, the maximum amplitude of Kerr signal is $\theta_{\rm K} \sim 100$ mdeg. The above results indicate that the observed spectral shape and amplitude of Kerr signals are quantitatively well reproduced by the DFT calculation with $M=0$. In Supplementary Fig. 3 and Supplementary Note III, we also performed similar DFT calculations for the easy-plane AFM state with and without tiny additional spin canting towards the $a$-axis, which confirmed that the calculated Kerr rotation $\theta_{\rm K}$ and Kerr ellipticity $\eta_{\rm K}$ is rarely affected by the small net magnetization $\Delta M_{\rm cant}$. The above results unambiguously prove that $\theta_{\rm K}$ in the present compound is dominantly induced by TtS-broken antiferromagnetic order itself, rather than net magnetization.

In Supplementary Note III and Supplementary Fig. 4, we also provided the theoretically calculated Kerr spectra for easy-axis AFM phase. It predicts the absence of spontaneous $\theta_{\rm K}$ and $\eta_{\rm K}$, in consistent with the experimental results (Fig. 3a) and symmetry-based analyses (Left panel of Fig. 3g).

In Fig. 4d, the Kerr rotation angle $|\theta_{\rm K}|$ is plotted against the spontaneous magnetization $\Delta M_{\rm cant}$ for $\alpha$-Fe$_2$O$_3$, as well as the other previously reported materials\cite{1969_Orthoferrites_Ultraviolet, 2013_MnGa_Tailoring, 2016_MnGa_Composition-tuned, 1992_MnGa_Ferromagnetic, 1995_MnPt3_Magneto-optical, 1994_MnBi_Kerr, 1991_MnSb_Magnetic, 1983_FerromagneticMaterials_Magneto-optical, 1987_PtMnSb_Magneto-optical, 1995_FePt_Perpendicular, 1981_Fe3O4_Kerr-effect, 1980_FerromagneticMaterials_Handbook, 1969_FerricOxide_Ultraviolet, 1981_Garnets_Magnetooptical, 1971_MagneticMaterials_Properties, 1965_Orthoferrites_Studies}, following Ref. \cite{Mn3Sn_Kerr}. Here, the conventional ferromagnets are located in the yellow-shaded region, where $|\theta_{\rm K}|$ almost linearly scales with $M$. In this case, the typical value of $|\theta_{\rm K}|/\Delta M_{\rm cant}$ ratio is $0.2 \sim 2.0$ deg~T$^{-1}$. On the other hand, the present antiferromagnetic $\alpha$-Fe$_2$O$_3$ clearly deviates from this trend, accompanied by a large Kerr rotation angle with a small net magnetization ($\Delta M_{\rm cant} \sim 0.006$~$\mu_{\rm B}$/Fe). Consequently, the Kerr rotation angle per unit magnetization reaches $|\theta_{\rm K}|/\Delta M_{\rm cant} \approx 47$ deg~T$^{-1}$, which is the largest value ever reported and one or two orders of magnitude larger than that of typical FMs. This further supports the nontrivial origin of spontaneous Kerr effect in $\alpha$-Fe$_2$O$_3$, and demonstrates that the TtS-broken collinear antiferromagnets can be a promising material platform to realize giant magneto-optical effect with vanishingly small net magnetization.

In this study, we reported a giant spontaneous magneto-optical Kerr effect in a room-temperature antiferromagnetic insulator $\alpha$-Fe$_2$O$_3$. Our first-principles calculations successfully reproduce both the spectral shape and absolute magnitude of the Kerr rotation and ellipticity with remarkable accuracy. It unambiguously establishes that this effect originates from the intrinsic $\uparrow\downarrow$ antiferromagnetic order itself, rather than net magnetization. Previously, spontaneous Hall and Kerr effects with vanishingly small magnetization have been explored mainly in a limited class of TtS-broken non-collinear helimagnets, such as Mn$_3X$ ($X=$ Sn, Ge)\cite{Mn3Sn_Kerr, Mn3Sn_Nature, Mn3Ge} and Co$M_3$S$_6$ ($M=$ Ta, Nb)\cite{CoNb3S6_transport, CoTa3S6_Seki, CoTa3S6_Batista}. Our results demonstrate that even the simplest collinear antiferromagnetic order can induce a giant spontaneous Kerr effect at room temperature, when combined with an appropriate crystal lattice symmetry.

The observed contrasting behaviors between the easy-plane and easy-axis AFM phases, i.e., the presence and absence of spontaneous Kerr response, respectively, suggest that the breaking of $\mathcal{T}t$ symmetry is a necessary but not sufficient condition for the emergence of a fictitious magnetic field, highlighting the importance of additional symmetry breaking. Notably, even in the latter easy-axis AFM phase, the application of an external magnetic field induces a magneto-optical response that is significantly larger than what would be expected from a simple $M$-driven mechanism. This enhancement can be attributed to the field-induced spin component that is equivalent to that of the easy-plane AFM state. These observations indicate that magneto-optical responses in TtS-broken antiferromagnets can be significantly amplified beyond conventional $M$-based mechanisms, which provides a powerful means for highly sensitive optical readout of $\uparrow\downarrow$ and $\downarrow\uparrow$ spin states. In such systems, efficient optical writing of $\uparrow\downarrow$ and $\downarrow\uparrow$ spin states via inverse effects (such as the inverse Faraday effect) may be also expected. Our findings suggest that TtS-broken antiferromagnets potentially serve as a viable alternative to ferromagnets for optical information technologies, and further exploration of this new class of materials and their unique properties associated with fictitious magnetic field would be interesting subjects for future study.

\begin{figure}
\begin{center}
\includegraphics*[width=12.2cm]{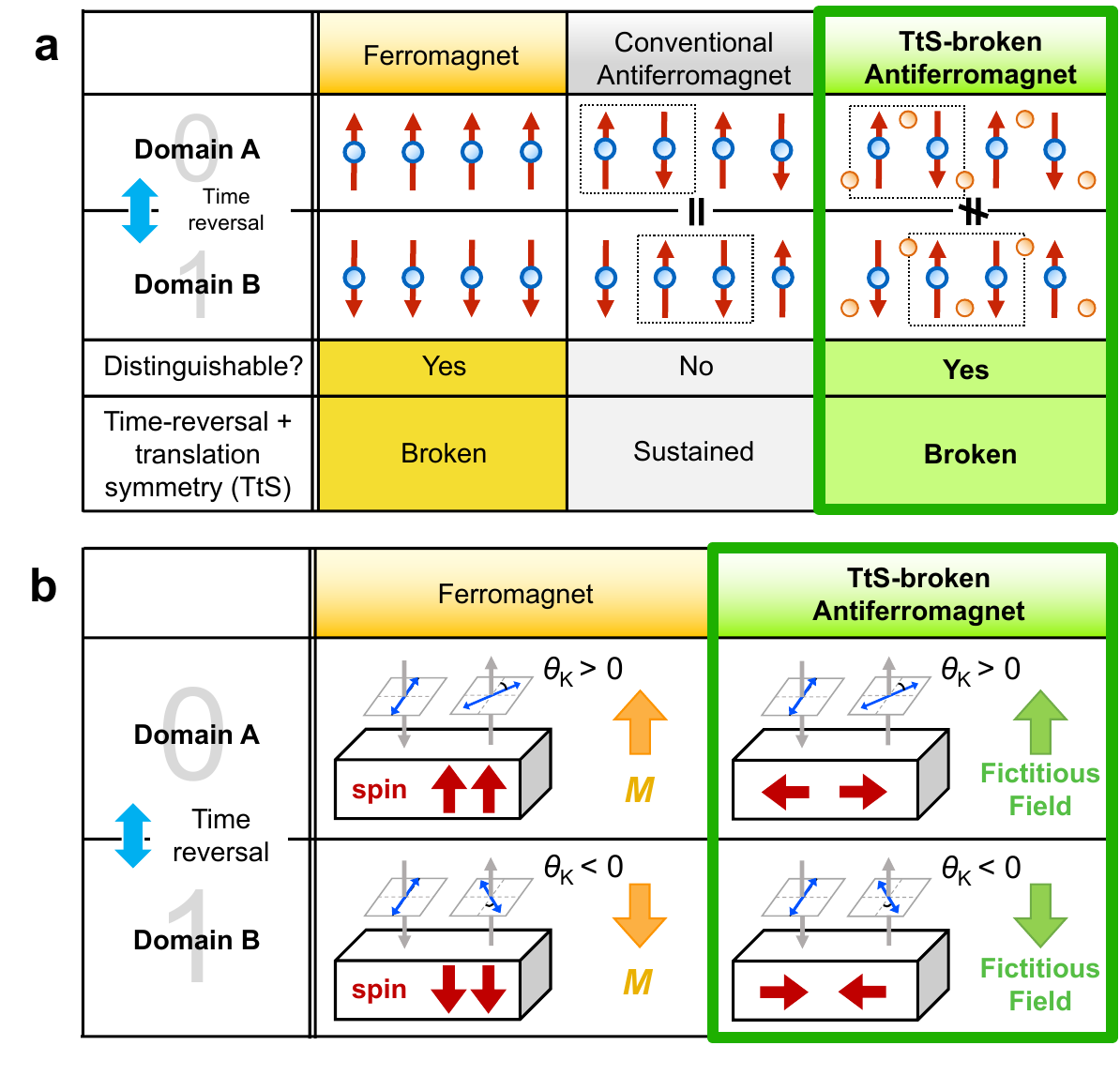}
\caption{{\bf Classification of various types of collinear magnets and spontaneous magneto-optical Kerr effect.} {\bf a}, Comparison of ferromagnet (FM), conventional antiferromagnet (AFM), and TtS-broken AFM. Red arrows represent local magnetic moments. Blue and red circles indicate magnetic and non-magnetic ions, respectively. Magnetic domains A and B are converted into each other by time-reversal operation $\mathcal{T}$. These two domains (i.e., $\uparrow \downarrow$ and  $\downarrow \uparrow$) are identical under a translation operation $t$ in conventional AFMs, but not identical in TtS-broken AFMs with appropriately located non-magnetic ions. {\bf b}, Comparison of spontaneous magneto-optical Kerr effect in FMs and TtS-broken AFMs. In the latter case, the collinear $\uparrow \downarrow$ spin order induces a fictitious magnetic field, which leads to the emergence of Kerr rotation even without magnetization $M$. Gray and blue arrows represent the propagation vector and polarization direction of the light, respectively. The sign of Kerr rotation angle $\theta_{\rm K}$ is opposite between the domains A and B. Note that the spontaneous Kerr effect is not allowed in conventional antiferromagnet, because of its TtS.}
\label{Fig1}
\end{center}
\end{figure}

\begin{figure}
\begin{center}
\includegraphics*[width=15.0cm]{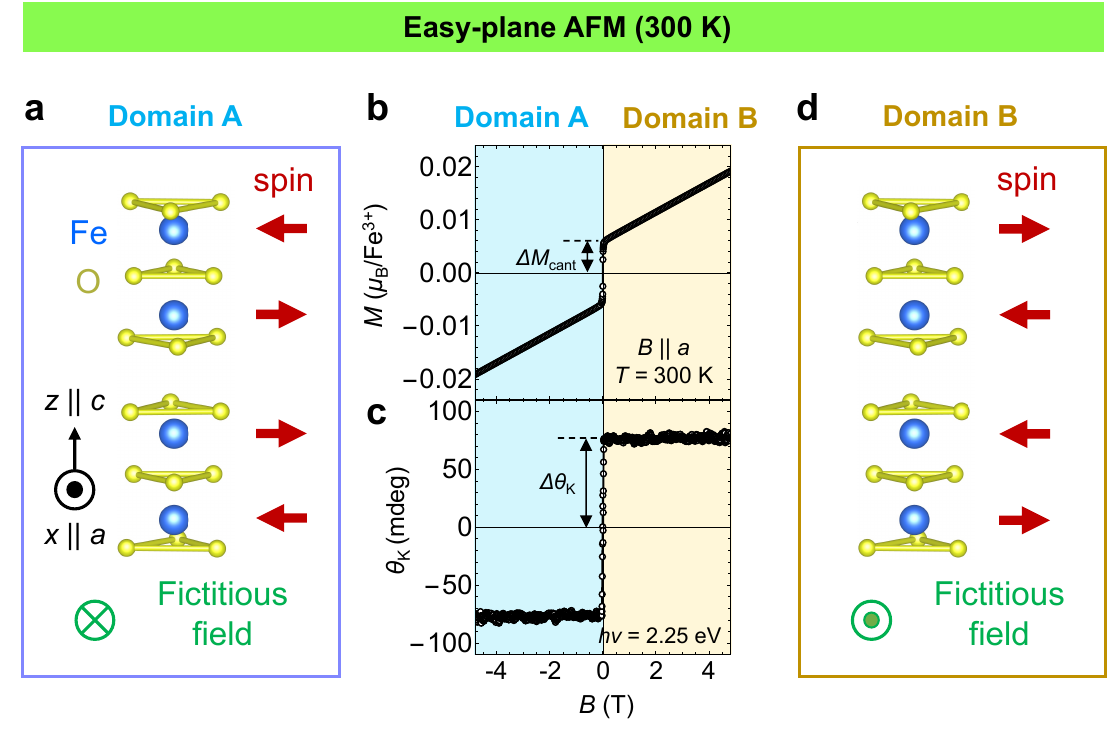}
\caption{{\bf Spontaneous magneto-optical Kerr effect in a collinear antiferromagnet $\alpha$-Fe$_2$O$_3$ at room temperature.} 
{\bf a,d}, Schematic illustration of the spin arrangements in the easy-plane AFM phase of $\alpha$-Fe$_2$O$_3$. Magnetic domains A ({\bf a}) and B ({\bf d}) (i.e., the $\uparrow \downarrow$ and $\downarrow \uparrow$ spin states) are converted into each other by time-reversal operation, which are characterized by the opposite sign of fictitious magnetic field and tiny spontaneous magnetization $\Delta M_{\rm cant}$ along the $a$-axis. {\bf b,c}, Magnetic field dependence of magnetization $M$ and polar magneto-optical Kerr rotation angle $\theta_{\rm K}$ measured with the normal incidence to the $a$-plane surface for $B \parallel a$ at 300 K in the easy-plane AFM phase. The incident light energy $h\nu$ is set as 2.25 eV. Here, the domain A (domain B) is selected by applying the negative (positive) sign of external magnetic field. The definition of $\Delta M_{\rm cant}$ and $\Delta \theta_{\rm K}$ are indicated in {\bf b} and {\bf c}, respectively.
}
\label{Fig2}
\end{center}
\end{figure}

\begin{figure}
\begin{center}
\includegraphics*[width=15cm]{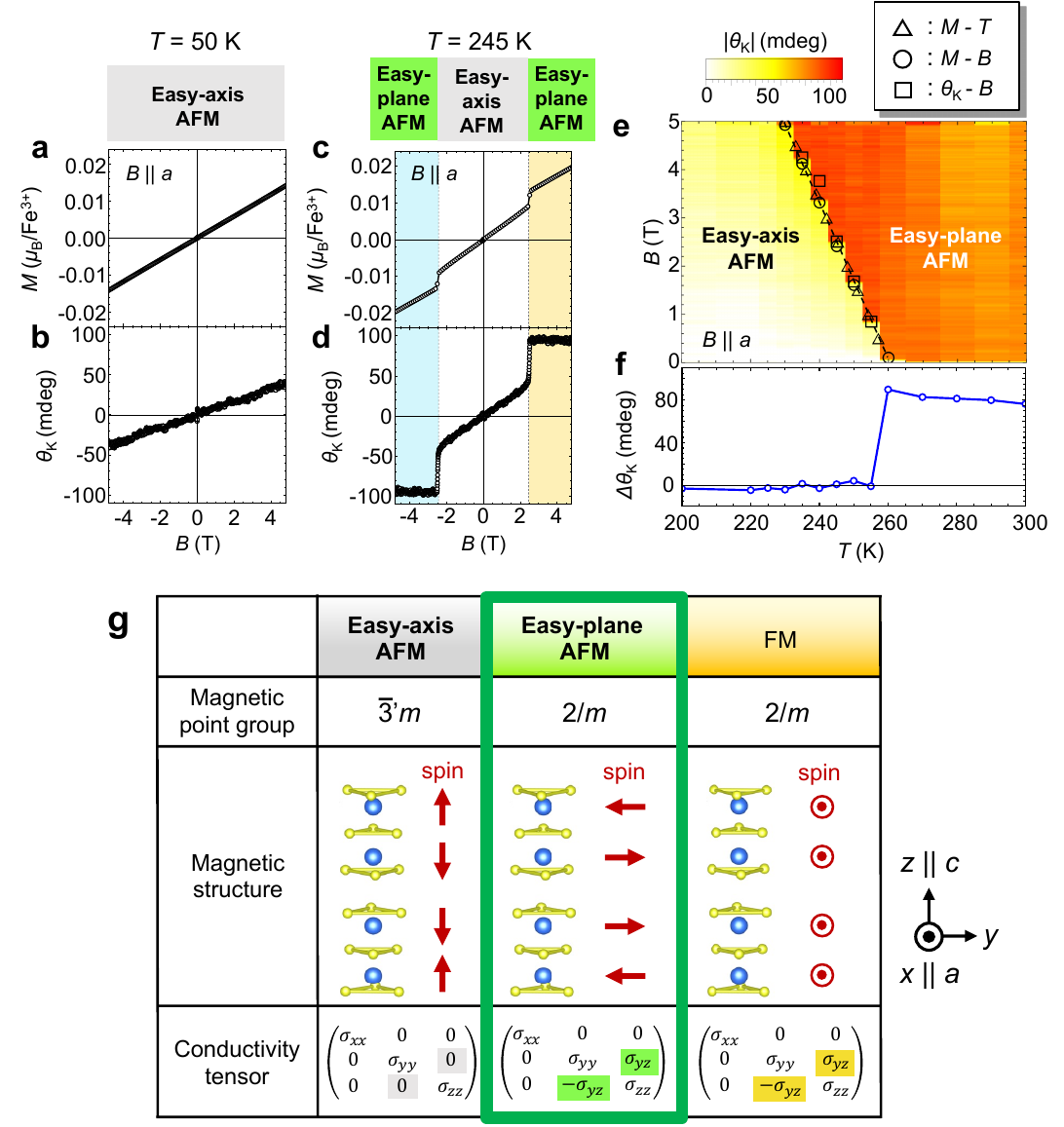}
\caption{{\bf The presence (absence) of spontaneous magneto-optical Kerr effect in the easy-plane (easy-axis) AFM state.} 
{\bf a-d}, Magnetic field dependence of magnetization $M$ and polar magneto-optical Kerr rotation angle $\theta_{\rm K}$ under $B \parallel a$ at 50 K ({\bf a,b}) and 245 K ({\bf c,d}). {\bf e}, $B$-$T$ magnetic phase diagram for $B \parallel a$, where the background color represents the amplitude of $\theta_{\rm K}$. {\bf f}, Temperature dependence of spontaneous Kerr rotation angle $\Delta\theta_{\rm K}$, obtained by linearly extrapolating the $\theta_{\rm K}$-$B$ curve to $B=0$ (as shown in Fig. 2c). {\bf g}, Symmetry analysis of various collinear spin arrangements on the $\alpha$-Fe$_2$O$_3$ crystal lattice. Magnetic point group, magnetic structure and corresponding symmetry-imposed shape of optical conductivity tensor $\tilde{\sigma}$ are presented for easy-axis AFM, easy-plane AFM and ferromagnetic (FM) spin states. }
\label{Fig3}
\end{center}
\end{figure}

\begin{figure}
\begin{center}
\includegraphics*[width=14cm]{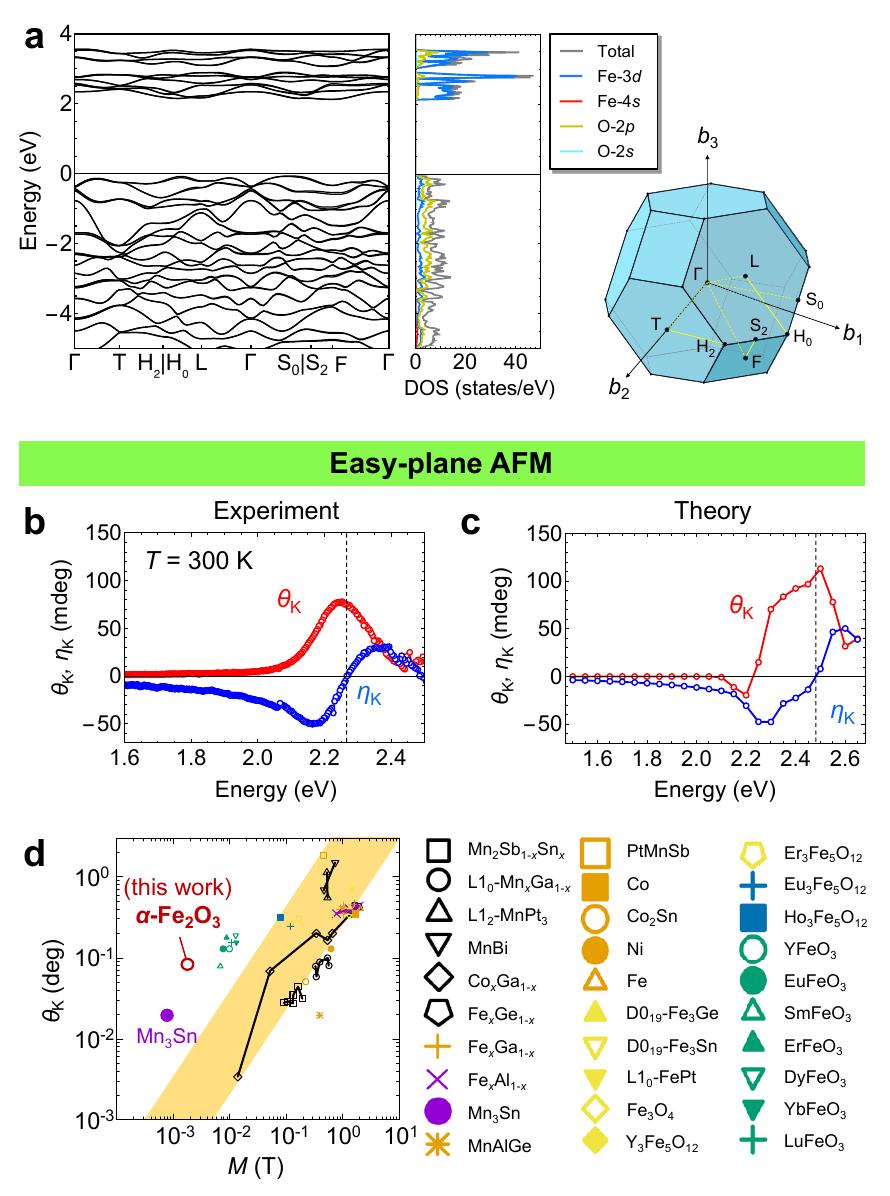}
\caption{{\bf Microscopic origin of spontaneous magneto-optical Kerr effect in the easy-plane AFM phase of $\alpha$-Fe$_2$O$_3$.} 
{\bf a}, Electronic band structure and density of state (DOS) of $\alpha$-Fe$_2$O$_3$ in the easy-plane AFM state, theoretically calculated based on DFT calculation with on-site Hubbard correlation $U$ = 4.0 eV. The letter symbols represent the positions in Brillouin zone, schematically illustrated in the right panel. {\bf b}, Experimental energy spectra of Kerr rotation angle $\theta_{\rm K}$ and Kerr ellipticity $\eta_{\rm K}$, measured at 300 K (easy-plane AFM state) with $B \parallel a$ at 1.0 T. {\bf c}, Theoretical energy spectra of $\theta_{\rm K}$ and $\eta_{\rm K}$ in the easy-plane AFM state with $M=0$, calculated based on the electronic structure in {\bf a}. {\bf d}, Full logarithmic plot of magnetization dependence of the polar magneto-optical Kerr rotation $\theta_{\rm K}$ for various magnetic materials. Conventional ferromagnets are located in the yellow shadowed region.}
\label{Fig4}
\end{center}
\end{figure}

\section*{Methods}

\subsection{Sample preparation.}

Plate-shaped single crystals of $\alpha$-Fe$_2$O$_3$, with flat surfaces normal to the $a$-axis polished to an optical grade, were purchased from Crystal Base Co., Ltd. The purity and single crystalline character of the sample were confirmed by the magnetization measurements and X-ray Laue back-reflection method, the latter of which ensures that the flat surfaces indeed correspond to the $a$-plane.

\subsection*{Magnetization measurement.}
Magnetization measurements were performed using a Magnetic Properties Measurement System (MPMS, Quantum Design) with a superconducting quantum interference device (SQUID) magnetometer on a rectangular-shaped sample.

\subsection*{Magneto-optical Kerr measurement.}

Polar Kerr measurements were performed by a standard method based on photoelastic optical phase modulator (PEM), where the incident and reflected lights are normal to the $a$-plane surface of the sample. A supercontinuum laser (SC-OEM, YSL Photonics) was used as a light source. The beam with the selective wavelength was extracted by using the wavelength selection module based on the acousto-optic crystal (AOTF-PRO, YSL Photonics). The beam was linearly polarized by using a Glan-laser prism and modulated for $\sim 50$ kHz and $\sim 227$ Hz with PEM and a mechanical chopper, respectively. The reflected beam then passed through a Glan-laser prism and detected with a Silicon photodetector. We performed synchronous detection of the reflected beam at the first and second harmonic frequency of the PEM modulation by using a lock-in amplifier. The static reflectivity was simultaneously measured by using a lock-in amplifier with the frequency of the mechanical chopper. External magnetic field up to 5 T was applied perpendicular to the sample surface, by using a superconducting magnet.

Since the Kerr signal is an odd function of magnetic field $B$, the obtained $\theta_{\rm K}-B$ and $\eta_{\rm K}-B$ profiles, as well as the energy spectra of the $\theta_{\rm K}$ and $\eta_{\rm K}$, are anti-symmetrized with respect to $B$ as $\theta_{\rm K}(+|B|)=-\theta_{\rm K}(-|B|)=(\theta^{\rm raw}_{\rm K}(+|B|)-\theta^{\rm raw}_{\rm K}(-|B|))/2$ and $\eta_{\rm K}(+|B|)=-\eta_{\rm K}(-|B|)=(\eta^{\rm raw}_{\rm K}(+|B|)-\eta^{\rm raw}_{\rm K}(-|B|))/2$, respectively, with $\theta^{\rm raw}_{\rm K}$ and $\eta^{\rm raw}_{\rm K}$ representing the raw experimental data. The measurement at 300 K is performed without an optical window. To eliminate the extrinsic contribution, a silver thin film was deposited on the half of sample surface by using the radio-frequency sputtering system, and the reference data of the silver surface was subtracted from the data of the $\alpha$-Fe$_2$O$_3$ surface. For the measurement at low temperature, $B$-induced Faraday rotation by an optical window was evaluated in advance, and subtracted from the data.

\subsection{First-principles calculations}
The electronic band structure and Kerr spectra of $\alpha$-Fe$_2$O$_3$ were calculated within spin DFT based on the projector augmented wave method~\cite{Kresse_PRB_1999}, as implemented in the Vienna Ab initio Simulation Package (VASP)~\cite{Kresse_PRB_1993, Kresse_PRB_1994}.
The calculations have been performed for the rhombohedral $R\bar{3}c$ structure, using the lattice constants $a = 5.427~\text{\AA}$ and $\alpha = 55.3^{\circ}$~\cite{Tamirat_Nanoscale_2016}.
In the easy-plane (easy-axis) antiferromagnetic state, a collinear antiferromagnetic order was assumed, with spins lying in the $c$-plane ($c$-axis).
The exchange-correlation functional was treated within the generalized gradient approximation (GGA) in the form proposed by Perdew-Burke-Ernzerhof~\cite{Perdew_PRL_1996}.
A plane-wave energy cutoff of 520 eV was employed, and the Brillouin zone was sampled using a $7 \times 7 \times 7$ $k$ grid.
To account for strong electron correlation effects in the Fe $3d$ orbitals, the GGA+U method was used with an effective on-site Hubbard interaction of $U = 4.0$ eV.
Maximally localized Wannier functions were constructed using the Wannier90 package~\cite{Mostofi_w90_2008, Pizzi_w90_2020} based on the spin DFT calculations, from which a 76-orbital Wannier tight-binding model was built using 256 Bloch bands on the $7 \times 7 \times 7$ $k$ grid, with the Fe-$d$ and O-$p$ orbitals taken as initial projections.
The Kerr rotation angle $\theta_{\rm K}$ and Kerr ellipticity $\eta_{\rm K}$ shown in Fig.~4(c) and Supplementary Figs.~3(b)-3(e) were calculated within linear response theory using the Kubo formula~\cite{Guo_PRB_1994, Guo_PRB_1995},
\begin{equation}
\theta_{\mathrm{K}} + i \eta_{\mathrm{K}} =
\frac{-\sigma_{yz}}{\sigma_0 \sqrt{1 + i (4\pi / \omega) \sigma_0}},
\qquad
\sigma_0 = \frac{\sigma_{yy} + \sigma_{zz}}{2},
\end{equation}
where the optical conductivity tensor was evaluated using the Wannier tight-binding Hamiltonian on a $30 \times 30 \times 30$ $k$-point grid.

\section*{Data availability}

The data presented in the current study are available from the corresponding authors on reasonable request.

\section*{Author contributions}

S.S. and R.T. conceived the project. H.Y., K.Y. and R.T. characterized the magnetic properties with the assistance of N.D.K, A.K., and S.S. Magneto-optical Kerr measurements were carried out by H.Y. and K.Y. First-principles calculation was performed by R.O., T.N., and R.A. The manuscript was written by H.Y. and S.S. with the assistance of R.O. and R.A. All the authors discussed the results and commented on the manuscript.

\section*{Acknowledgements} The authors thank N. Ogawa and R. Hirakida and for enlightening discussions and experimental helps. This work was partly supported by Grants-In-Aid for Scientific Research (grant nos 21H04440, 21H04990, 21K13876, 21K18595, 22KJ1061, 22H04965, 24K00579, 24H02235, 24K00581, 25K21684, 25H00420, 25H00611, 25H01252, 25H01246, 25K00956) from JSPS, PRESTO (grant nos JPMJPR18L5, JPMJPR20B4) and CREST (grant no. JPMJCR23O4) from JST, Katsu Research Encouragement Award and UTEC-UTokyo FSI Research Grant Program of the University of Tokyo, Special Postdoctoral Researcher Program at RIKEN, Asahi Glass Foundation and Murata Science Foundation. H.Y. was supported by JSPS through the research fellowship for young scientists and Iwadare Scholarship Foundation. The illustrations of crystal structure were drawn by VESTA\cite{VESTA}.

\section*{Additional information}
Supplementary Information is available in the online version of the paper. Correspondence and requests for materials should be addressed to H.Y. and S.S.

\section*{Competing financial interests}
The authors declare that they have no competing financial interests.

\clearpage

\end{document}

\title{Supplementary Materials:\\
Large magneto-optical Kerr effect induced by collinear antiferromagnetic order}

\author{H. Yoshimochi$^1$, K. Yoshida$^1$, R. Oiwa$^{2,3}$, T. Nomoto$^{4}$, N. D. Khanh$^{1}$, A. Kitaori$^{1,5,6}$, R. Takagi$^{5,7}$, R. Arita$^{2, 8}$, S. Seki$^{1,6,9}$}

\affiliation{$^1$ Department of Applied Physics and Quantum-Phase Electronics Center (QPEC), University of Tokyo, Tokyo 113-8656, Japan, \\ 
$^2$ RIKEN Center for Emergent Matter Science (CEMS), Wako 351-0198, Japan,\\
$^3$ Department of Physics, Hokkaido University, Sapporo 060-0810, Japan, \\
$^4$ Department of Physics, Tokyo Metropolitan University, Hachioji, Tokyo 192-0397, Japan,\\
$^5$ PRESTO, Japan Science and Technology Agency (JST), Kawaguchi 332-0012, Japan, \\ 
$^6$ Research Center for Advanced Science and Technology (RCAST), University of Tokyo, Tokyo 153-8904, Japan, \\ 
$^7$ Institute for Solid State Physics, University of Tokyo, Kashiwa 277-8581, Japan,\\
$^8$ Department of Physics, University of Tokyo, Tokyo 113-8656, Japan, \\ 
$^9$ Institute of Engineering Innovation, University of Tokyo, Tokyo 113-0032, Japan}

\maketitle

\section{Magneto-optical Kerr ellipticity}

\begin{figure}[b]
\begin{center}
\includegraphics*[width=15.5cm]{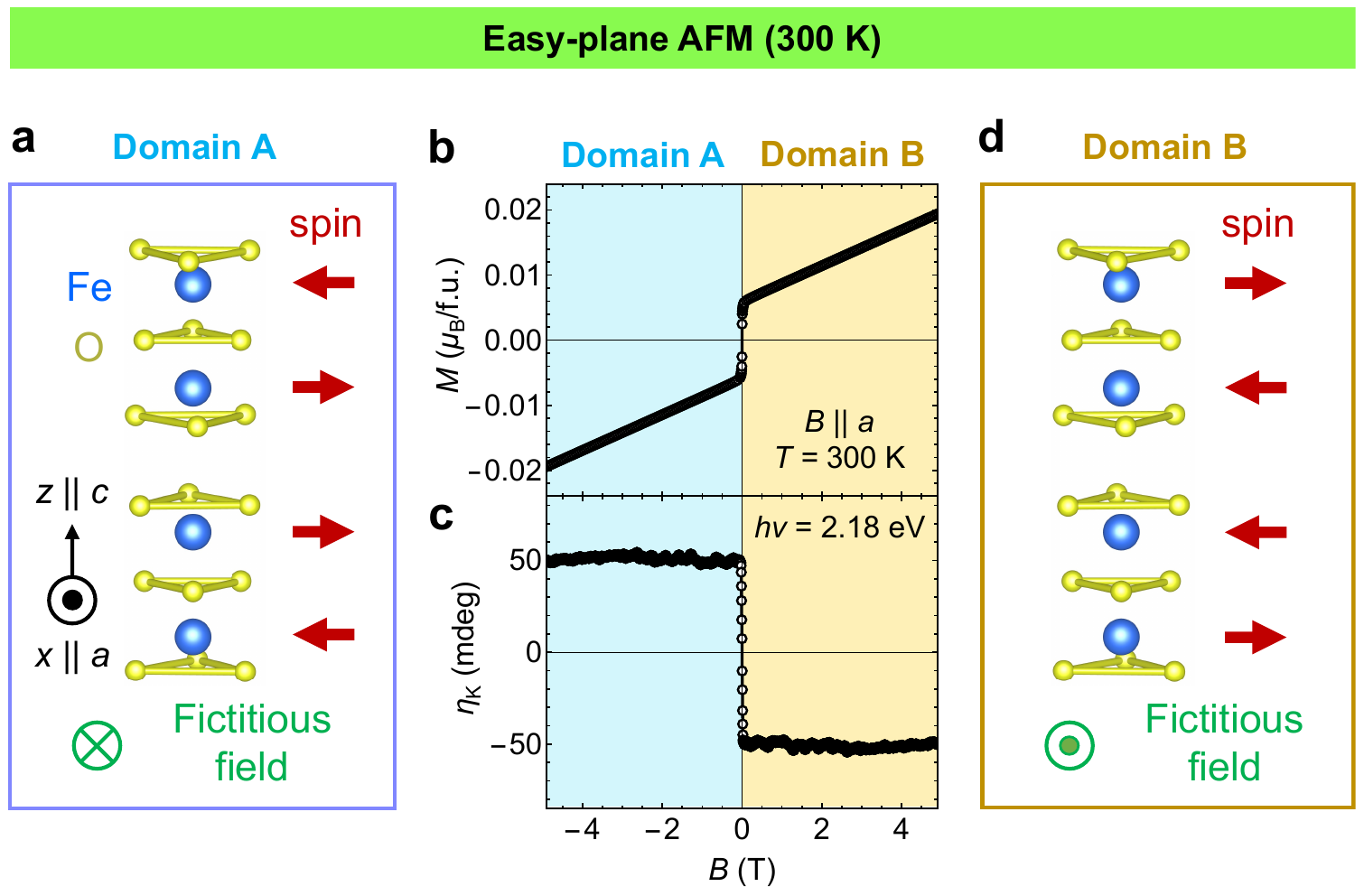}
\caption{{\bf Magnetic-field-dependence of magnetization and magneto-optical Kerr ellipticity in the easy-plane AFM state.} 
{\bf a,d}, Schematic illustration of the spin arrangements in the easy-plane AFM phase of $\alpha$-Fe$_2$O$_3$. Magnetic domains A ({\bf a}) and B ({\bf d}) (i.e., the $\uparrow \downarrow$ and $\downarrow \uparrow$ spin states) are converted into each other by time-reversal operation, which are characterized by the opposite sign of fictitious magnetic field and tiny spontaneous magnetization $\Delta M_{\rm cant}$ along the $a$-axis. {\bf b,c}, Magnetic field dependence of magnetization $M$ and polar magneto-optical Kerr ellipticity $\eta_{\rm K}$ measured with the normal incidence to the $a$-plane surface for $B \parallel a$ at 300 K in the easy-plane AFM phase. The incident light energy $h\nu$ is set as 2.18 eV.
}
\label{FigS1}
\end{center}
\end{figure}

In this section, the measurement on magneto-optical Kerr ellipticity is discussed. Since the complex Kerr coefficient is defined as $\Phi_{\rm K} = \theta_{\rm K} + i\eta_{\rm K}$, observing both real and imaginary parts ensures the consistency of the optical response via the Kramers-Kronig relations\cite{KerrTextbook}. To discuss the magneto-optical response from a comprehensive perspective, we measured the magneto-optical Kerr ellipticity $\eta_{\rm K}$ in addition to the Kerr rotation $\theta_{\rm K}$ in the easy-plane AFM phase. Since the Kerr signal should be odd against $B$, the measured $\theta_{\rm K}$-$B$ and $\eta_{\rm K}$-$B$ profiles are anti-symmetrized with respect to $B$ to eliminate the possible extrinsic artifacts or the Voigt effect, as detailed in the Methods section. 

Supplementary Figs. \ref{FigS1}b and c show the magnetic field dependence of $M$ and $\eta_{\rm K}$ measured at room temperature ($T=300$ K) for $B \parallel a$. The incident photon energy is set as $h\nu = 2.18$ eV, which corresponds to the peak position of $\eta_{\rm K}$ in the measured energy spectrum shown in Fig. 4b in the main text. The application of $B \parallel a$ leads to the deterministic switching of two distinct magnetic domains A and B (i.e., $\uparrow \downarrow$ and $\downarrow \uparrow$ spin states), characterized by the opposite signs of tiny spontaneous magnetization $\Delta M_{\rm cant}$ and fictitious magnetic field along the $a$-axis, as shown in Supplementary Figs. \ref{FigS1}a and d. Corresponding to the domain switching, the Kerr ellipticity exhibits a clear step-like anomaly, which evidences the presence of a large amplitude of spontaneous $\eta_{\rm K}$ around $B = 0$. While $M$ shows $B$-linear increase up to 5 T, $\eta_{\rm K}$ is almost constant against $B$. It indicates that the observed large spontaneous $\eta_{\rm K}$ originates from antiferromagnetic order itself, rather than tiny net magnetization. As shown in Fig. 4 in the main text, the energy spectrum of $\eta_{\rm K}$ exhibits a dispersive structure at the energy corresponding to the peak of $\theta_{\rm K}$, which is in agreement with the Kramers-Kronig relations. These results confirm the validity of the Kerr signal observed in this experiment.

\section{Magnetic field dependence of magnetization and magneto-optical responses at various temperatures}

\begin{figure}[t]
\begin{center}
\includegraphics*[width=15cm]{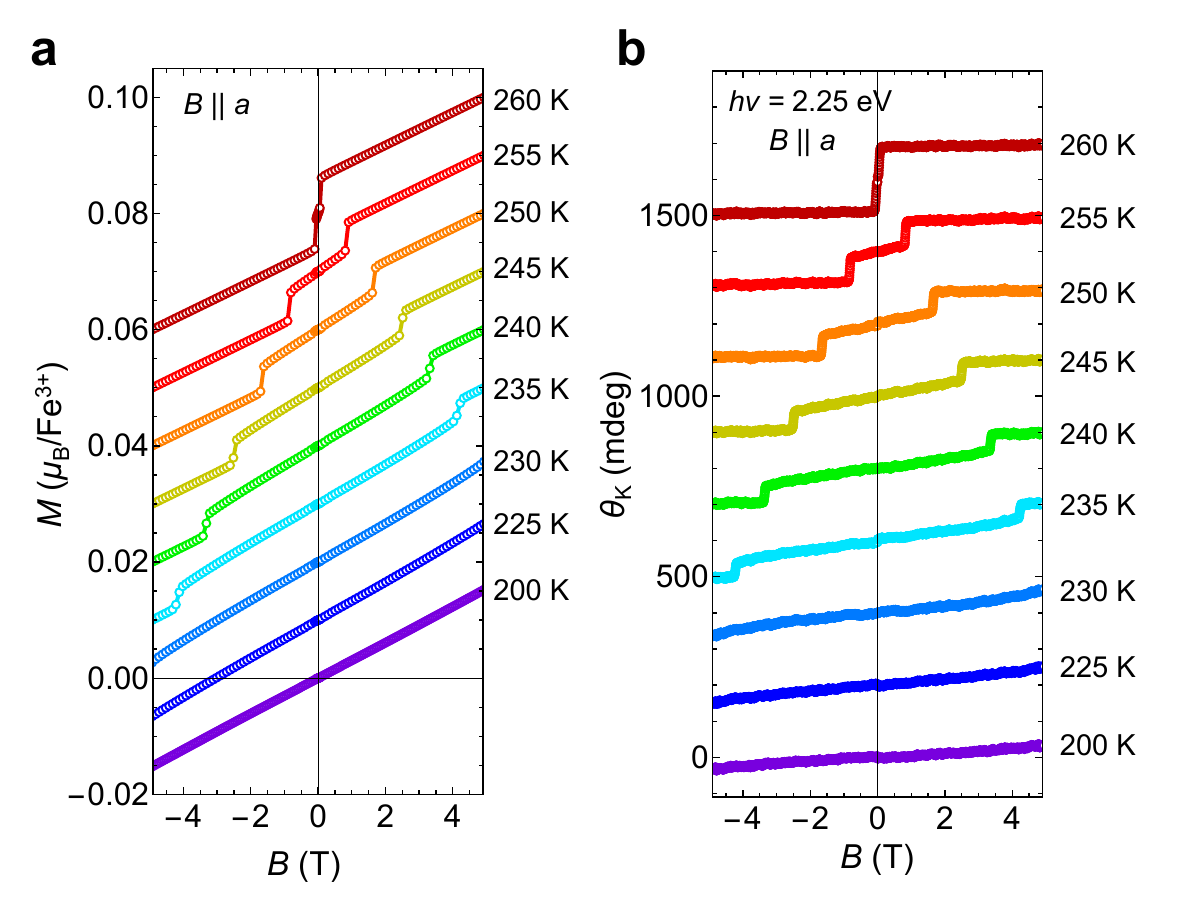}
\caption{{\bf Magnetization and polar magneto-optical Kerr rotation profiles of $\alpha$-Fe$_2$O$_3$ measured at various temperature.} {\bf a}, Magnetic-field dependence of magnetization $M$ and polar magneto-optical Kerr rotation $\theta_{\rm K}$ measured for $B \parallel a$ with 2.25 eV at various temperatures. Each data is arbitrarily shifted along the vertical direction for clarity.
}
\label{FigS2}
\end{center}
\end{figure}

Supplementary Figs. \ref{FigS2}a and b show the magnetic-field-dependence of the magnetization $M$ and the polar magneto-optical Kerr rotation angle $\theta_{\rm K}$ measured at various temperatures. The measurements were performed in the polar Kerr configuration with $B \parallel a$, where the incident and reflected light are normal to the $a$-plane sample surface. The photon energy of the incident light was fixed at $h\nu = 2.25$ eV, which corresponds to the peak energy of the Kerr rotation spectrum. 

As discussed in the main text, this compound undergoes a Morin transition at $T_{\rm M} \approx 260$ K. In the easy-plane AFM state above $T_{\rm M}$, step-like anomalies associated with spontaneous magnetization and Kerr rotation are identified at around $B=0$. In this state, while $M$ linearly increases as a function of $B$, $\theta_{\rm K}$ remains almost constant against $M$. In the easy-axis AFM state, on the other hand, spontaneous magnetization and Kerr rotation disappear, and both $M$ and $\theta_{\rm K}$ linearly increase as a function of $B$. The latter feature is probably due to the $B$-induced tilting of N\'{e}el vector from the $c$-axis toward the $y$-axis (i.e., the direction normal to both $a$- and $c$-axes), which induces the same spin component as the easy-plane AFM state. The observed presence/absence of spontaneous Kerr signal in each phase is fully consistent with the symmetry analysis in Fig. 3g in the main text.

Between 235 K and 260 K at around the Morin transition temperature, the application of $B \parallel a$ induces the transition from easy-axis AFM phase to easy-plane AFM phase, which is identified as a step-like anomalies in both $M$ and $\theta_{\rm K}$ profiles. These data are used to plot $\theta_{\rm K}$ as a function of $B$ and $T$ in Fig. 3e in the main text.

\section{First-principles calculation of magneto-optical Kerr responses}

\begin{figure}[p]
\begin{center}
\includegraphics*[width=13cm]{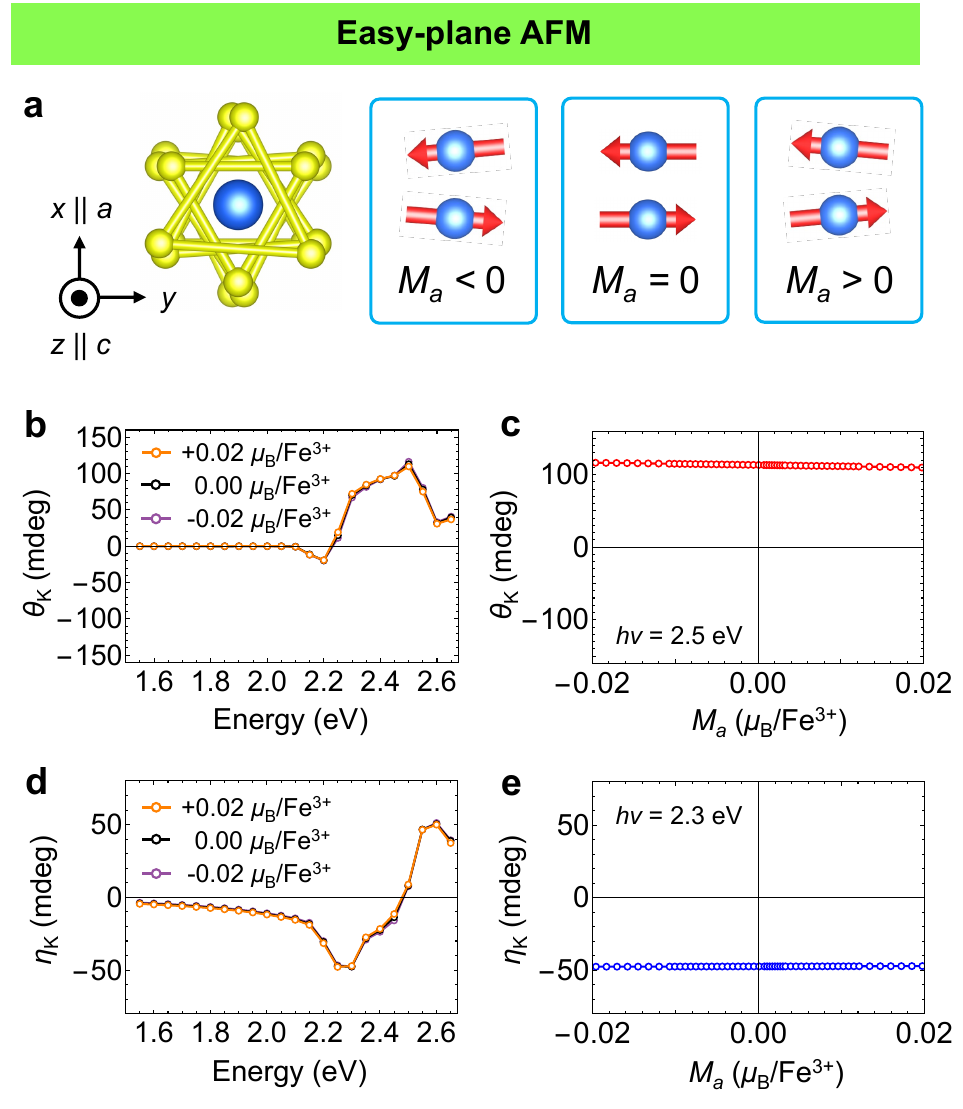}
\caption{{\bf First principles calculation of magneto-optical Kerr rotation and ellipticity in the easy-plane AFM state of $\alpha$-Fe$_2$O$_3$ with and without spin canting.} {\bf a}, Schematic illustration of the tiny spin canting along $a$-axis. The right panel represents a pair of spins extracted from the four Fe$^{3+}$ spins stacking along the $c$-axis within a unit cell.
{\bf b,d}, Energy spectrum of Kerr rotation $\theta_{\rm K}$ ({\bf b}) and Kerr ellipticity $\eta_{\rm K}$ ({\bf d}) calculated with various spin canting along $x \parallel a$ direction. The assumed magnetization value $M_a = 0.02 \mu_{\rm B}/{\rm Fe}^{3+}$ corresponds to the experimental $M_a$ value induced by applying strong external magnetic field as large as 5 T, as shown in Fig. 2b in the main text. 
{\bf c,e}, $\theta_{\rm K}$ ({\bf c}) and $\eta_{\rm K}$ ({\bf e}) as a function of $M_a$, calculated at $h\nu = 2.5$ eV and $h\nu = 2.3$ eV, respectively.
}
\label{FigS3}
\end{center}
\end{figure}

\begin{figure}[t]
\begin{center}
\includegraphics*[width=13cm]{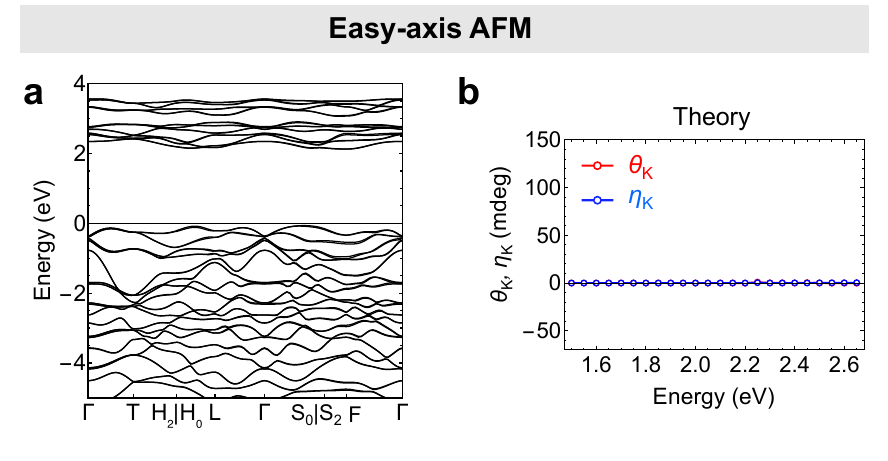}
\caption{{\bf First principles calculation of electronic band structure and magneto-optical Kerr responses in the easy-axis AFM state of $\alpha$-Fe$_2$O$_3$.} {\bf a}, Electronic band structure of $\alpha$-Fe$_2$O$_3$ in the easy-axis AFM state, theoretically estimated based on DFT calculation with on-site Hubbard correlation $U$ = 4.0 eV. The letter symbols represent the positions in Brillouin zone, schematically illustrated in Fig. 4a in the main text. {\bf b}, Theoretical energy spectra of $\theta_{\rm K}$ and $\eta_{\rm K}$ in the easy-axis AFM state with $M=0$, calculated based on the electronic structure in {\bf a}.
}
\label{FigS4}
\end{center}
\end{figure}

In this section, the microscopic mechanism of the Kerr signal in $\alpha$-Fe$_2$O$_3$ is discussed. A central claim of this study is that the observed giant Kerr signal originates from the TtS-broken collinear antiferromagnetic order, rather than tiny net magnetization caused by spin canting. To verify this point, we performed the first-principles calculation based on density functional theory (DFT) for the easy-plane antiferromagnetic state with and without considering additional spin canting. (See the main text for the detailed discussion on the calculated band structure.)

In the following, we assume the domain A with antiparallel spins along the $y$-axis. The additional tilting of spins from the $y$-axis toward the $a$-axis is introduced, which results in a finite net magnetization $M$ along the $a$-axis ($M_a$). Supplementary Fig. \ref{FigS3}a schematically illustrates the assumed spin configuration, where the antiparallel spin pair shown on the right panel represents two Fe spins extracted from the four spins stacked along the $c$-axis within a magnetic unit cell. Supplementary Figs. \ref{FigS3}b and d present the calculated energy spectra of $\theta_{\rm K}$ and $\eta_{\rm K}$ for three distinct net magnetization values: $M_a = -0.02$, 0.00, and 0.02 $\mu_{\rm B}/{\rm Fe}^{3+}$, respectively. Note that the value of $0.02 \mu_{\rm B}/{\rm Fe}^{3+}$ is comparable to the experimental magnetization observed under a strong external magnetic field of 5 T. Remarkably, the calculated spectra of $\theta_{\rm K}$ and $\eta_{\rm K}$ remain almost unchanged regardless of the presence/absence or sign of $M_a$ value, which clearly demonstrate that the observed giant Kerr signal originates from TtS-broken AFM order itself, rather than net magnetization.

Supplementary Figs. \ref{FigS3}c and e present the calculated values of Kerr rotation angle $\theta_{\rm K}$ at 2.5 eV and Kerr ellipticity $\eta_{\rm K}$ at 2.3 eV as a function of $M_a$. The calculated $\theta_{\rm K}$ and $\eta_{\rm K}$ remain nearly constant regardless of the $M_a$ value. Moreover, $\theta_{\rm K}$ and $\eta_{\rm K}$ retain large finite values even when $M_a$ is strictly zero. The above results indicate that net magnetization is not relevant for the observed giant magneto-optical Kerr effect in the present compound, and the contribution of TtS-broken AFM order is dominant.

In Supplementary Fig. 4, we also presented the theoretically calculated electronic band structure and energy spectra of $\theta_{\rm K}$ and $\eta_{\rm K}$ for the easy-axis antiferromagnetic state. It predicts the absence of spontaneous $\theta_{\rm K}$ and $\eta_{\rm K}$, which is consistent with the experimental results (Fig. 3a in the main text) and symmetry-based analyses (Left panel of Fig. 3g in the main text).

\section{Magnetic domains in the easy-plane antiferromagnetic phase}

\begin{figure}[b]
\begin{center}
\includegraphics*[width=13cm]{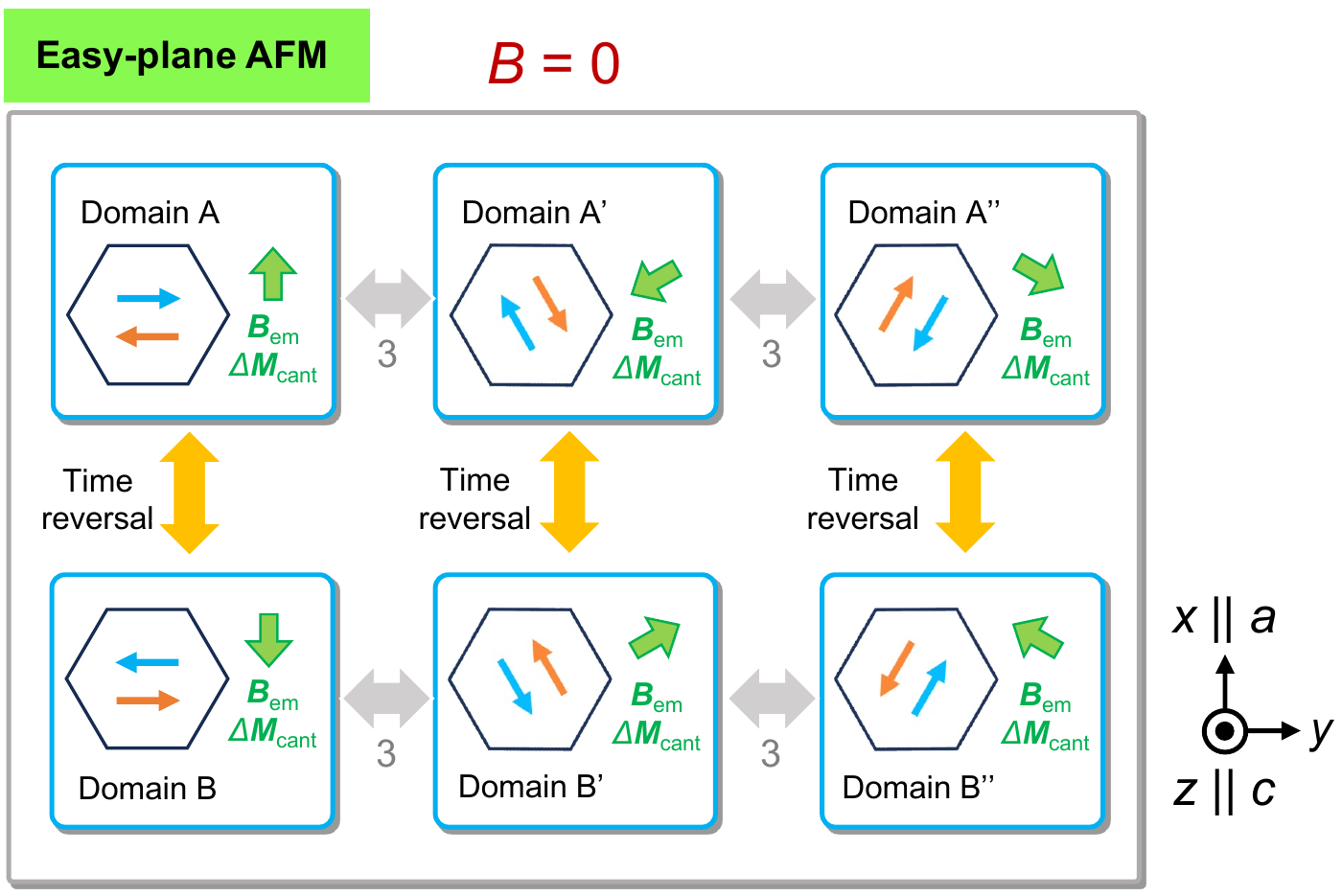}
\caption{{\bf Magnetic domains in the easy-plane antiferromagnetic state.} Schematic illustration of the six equivalent magnetic domains in the easy-plane antiferromagnetic state, where red (blue) arrows represent the local magnetic moment of the first and the fourth (the second and the third) Fe layers, respectively. These magnetic domains are converted into each other by time-reversal operation or three-fold rotation. Green arrows represent the orientation of the fictitious magnetic field ${\bf B}_{\rm em}$ and tiny spontaneous magnetization $\Delta M_{\rm cant}$. 
}
\label{FigS5}
\end{center}
\end{figure}

In this section, we discuss the magnetic domains in the easy-plane AFM phase of $\alpha$-Fe$_2$O$_3$. When the magnetic order breaks the symmetry of original crystal structure, there should appear multiple equivalent magnetic domains that are converted into each other by the broken symmetry elements. In case of $\alpha$-Fe$_2$O$_3$, the easy-plane antiferromagnetic order breaks the three-fold rotational symmetry and time-reversal symmetry, and it leads to the appearance of six possible antiferromagnetic domains (i.e. the domains A, B, A', B', A'', and B'') as shown in in Supplementary Fig. 5.

In the domain A, for example, antiparallel spins are oriented along the $y$-axis, whose magnetic symmetry allows the emergence of fictitious magnetic field ${\bf B}_{\rm em}$ along the $a$-axis as discussed in Fig. 3g in the main text. In Supplementary Fig. 5, the direction of ${\bf B}_{\rm em}$ for each domain is shown by the green bold arrow. Here, the domains A, A' and A'' are converted into each other by three-fold rotation. The domains B, B' and B'' are related with the domains A, A' and A'' by the time-reversal operation, respectively, and therefore characterized by the opposite direction of ${\bf B}_{\rm em}$. Since these domains host tiny spontaneous magnetization ${\bf \Delta M}_{\rm cant}$ along the direction of fictitious magnetic field ${\bf B}_{\rm em}$, the application of external magnetic field ${\bf B}$ leads to the selection of the domain with ${\bf \Delta M}_{\rm cant} \parallel {\bf B}$. For example, the application of positive (negative) sign of ${\bf B} \parallel a$ should lead to the selection of domains A (domain B) characterized by the positive (negative) sign of spontaneous magnetization ${\bf \Delta M}_{\rm cant} \parallel a$. 

 At $B=0$, these six domains are expected to coexist with the equal population. To extract the amplitude of spontaneous magnetization $\Delta M_{\rm cant}$ and spontaneous Kerr rotation angle $\Delta \theta_{\rm K}$ originating from a specific domain, $M$-$B$ and $\theta_{\rm K}$-$B$ curves are linearly extrapolated to $B=0$. The estimated $\Delta \theta_{\rm K}$ value is plotted in Figs. 3e and f in the main text.

\newpage